


\font\twelverm=cmr10  scaled 1200   \font\twelvei=cmmi10  scaled 1200
\font\twelvesy=cmsy10 scaled 1200   \font\twelveex=cmex10 scaled 1200
\font\twelvebf=cmbx10 scaled 1200   \font\twelvesl=cmsl10 scaled 1200
\font\twelvett=cmtt10 scaled 1200   \font\twelveit=cmti10 scaled 1200
\font\twelvesc=cmcsc10 scaled 1200
\skewchar\twelvei='177   \skewchar\twelvesy='60


\def\twelvepoint{\normalbaselineskip=12.4pt plus 0.1pt minus 0.1pt
  \abovedisplayskip 12.4pt plus 3pt minus 9pt
  \belowdisplayskip 12.4pt plus 3pt minus 9pt
  \abovedisplayshortskip 0pt plus 3pt
  \belowdisplayshortskip 7.2pt plus 3pt minus 4pt
  \smallskipamount=3.6pt plus1.2pt minus1.2pt
  \medskipamount=7.2pt plus2.4pt minus2.4pt
  \bigskipamount=14.4pt plus4.8pt minus4.8pt
  \def\rm{\fam0\twelverm}          \def\it{\fam\itfam\twelveit}%
  \def\sl{\fam\slfam\twelvesl}     \def\bf{\fam\bffam\twelvebf}%
  \def\mit{\fam 1}                 \def\cal{\fam 2}%
  \def\sc{\twelvesc}               \def\tt{\twelvett}
  \def\sf{\twelvesf}
  \textfont0=\twelverm   \scriptfont0=\tenrm   \scriptscriptfont0=\sevenrm
  \textfont1=\twelvei    \scriptfont1=\teni    \scriptscriptfont1=\seveni
  \textfont2=\twelvesy   \scriptfont2=\tensy   \scriptscriptfont2=\sevensy
  \textfont3=\twelveex   \scriptfont3=\twelveex  \scriptscriptfont3=\twelveex
  \textfont\itfam=\twelveit
  \textfont\slfam=\twelvesl
  \textfont\bffam=\twelvebf \scriptfont\bffam=\tenbf
  \scriptscriptfont\bffam=\sevenbf
  \normalbaselines\rm}



\def\beginlinemode{\endmode
  \begingroup\parskip=0pt \obeylines\def\\{\par}\def\endmode{\par\endgroup}}
\def\beginparmode{\endmode
  \begingroup \def\endmode{\par\endgroup}}
\let\endmode=\par
{\obeylines\gdef\
{}}
\def\singlespace{\baselineskip=\normalbaselineskip}
\def\oneandathirdspace{\baselineskip=\normalbaselineskip
  \multiply\baselineskip by 4 \divide\baselineskip by 3}
\def\oneandahalfspace{\baselineskip=\normalbaselineskip
  \multiply\baselineskip by 3 \divide\baselineskip by 2}
\def\doublespace{\baselineskip=\normalbaselineskip \multiply\baselineskip by 2}

\newcount\firstpageno
\firstpageno=2
\footline={\ifnum\pageno<\firstpageno{\hfil}\else{\hfil\twelverm\folio\hfil}\fi}

\def\toppageno{\global\footline={\hfil}\global\headline
  ={\ifnum\pageno<\firstpageno{\hfil}\else{\hfil\twelverm\folio\hfil}\fi}}
\let\rawfootnote=\footnote              
\def\footnote#1#2{{\rm\singlespace\parindent=0pt\parskip=0pt
  \rawfootnote{#1}{#2\hfill\vrule height 0pt depth 6pt width 0pt}}}
\def\raggedcenter{\leftskip=4em plus 12em \rightskip=\leftskip
  \parindent=0pt \parfillskip=0pt \spaceskip=.3333em \xspaceskip=.5em
  \pretolerance=9999 \tolerance=9999
  \hyphenpenalty=9999 \exhyphenpenalty=9999 }
\def\dateline{\rightline{\ifcase\month\or
  January\or February\or March\or April\or May\or June\or
  July\or August\or September\or October\or November\or December\fi
  \space\number\year}}
\def\received{\vskip 3pt plus 0.2fill
 \centerline{\sl (Received\space\ifcase\month\or
  January\or February\or March\or April\or May\or June\or
  July\or August\or September\or October\or November\or December\fi
  \qquad, \number\year)}}


\hsize=6.125truein
\hoffset=0.0truein
\vsize=9.0truein
\voffset=0.25truein
\parskip=\medskipamount
\nopagenumbers
\twelvepoint
\doublespace
\def\\{\cr}
\overfullrule=0pt 



\def\tutp#1{
  \rightline{\rm TUTP--#1}} 

\def\title#1{                   
   \null \vskip 3pt plus 0.3fill \beginlinemode
   \doublespace \raggedcenter {\bf #1} \vskip 3pt plus 0.1 fill}

\def\author                     
  {\vskip 3pt plus 0.1fill \beginlinemode \doublespace \raggedcenter}

\def\affil                      
  {\vskip 3pt \beginlinemode \doublespace \raggedcenter \it}

\def\abstract                   
  {\vskip 3pt plus 0.1fill \subhead {Abstract:}
   \beginparmode \narrower \oneandahalfspace }

\def\endtopmatter               
  {\vskip 3pt plus 0.1fill \endpage \body}

\def\body                       
  {\beginparmode}               

\def\head#1{                    
   \goodbreak \vskip 0.4truein  
  {\immediate\write16{#1} \raggedcenter {\sc #1} \par}
   \nobreak \vskip 3pt \nobreak}

\def\subhead#1{                 
  \vskip 0.25truein             
  {\raggedcenter {\it #1} \par} \nobreak \vskip 3pt \nobreak}

\def\beneathrel#1\under#2{\mathrel{\mathop{#2}\limits_{#1}}}

\def\refto#1{${\,}^{#1}$}       

\newdimen\refskip \refskip=0pt
\def\references         
  {\head{References}    
   \beginparmode \frenchspacing \parindent=0pt \leftskip=\refskip
   \parskip=0pt \everypar{\hangindent=20pt\hangafter=1}}

\gdef\refis#1{\item{#1.\ }}                     

\gdef\journal#1, #2, #3 {               
    {\it #1}, {\bf #2}, #3.}            




\def\endreferences{\body}

\def\figurecaptions             
  {\endpage \beginparmode \head{Figure Captions}
   \parskip=3pt \everypar{\hangindent=20pt\hangafter=1} }

\def\endpage                    
  {\vfill\eject}

\def\endpaper   {\endmode\vfill\supereject}
\def\endjnl     {\endpaper\end}


\def\ref#1{ref.{#1}}                    
\def\Ref#1{Ref.{#1}}                    
\def\[#1]{[\cite{#1}]}
\def\cite#1{{#1}}


\def\(#1){(\call{#1})}
\def\call#1{{#1}}
\def\frac#1#2{{#1 \over #2}}
\def\half{  {\frac 12}}

\def\12{{1\over2}}

\def\sla{\raise.15ex\hbox{$/$}\kern-.57em}
\def\leaderfill{\leaders\hbox to 1em{\hss.\hss}\hfill}
\def\twiddle{\lower.9ex\rlap{$\kern-.1em\scriptstyle\sim$}}
\def\bigtwiddle{\lower1.ex\rlap{$\sim$}}
\def\gtwid{\mathrel{\raise.3ex\hbox{$>$\kern-.75em\lower1ex\hbox{$\sim$}}}}
\def\ltwid{\mathrel{\raise.3ex\hbox{$<$\kern-.75em\lower1ex\hbox{$\sim$}}}}
\def\square{\kern1pt\vbox{\hrule height 1.2pt\hbox{\vrule width 1.2pt\hskip 3pt
   \vbox{\vskip 6pt}\hskip 3pt\vrule width 0.6pt}\hrule height 0.6pt}\kern1pt}
\def\tdot#1{\mathord{\mathop{#1}\limits^{\kern2pt\ldots}}}

\def\pmb#1{\setbox0=\hbox{#1}%
  \kern-.025em\copy0\kern-\wd0
  \kern  .05em\copy0\kern-\wd0
  \kern-.025em\raise.0433em\box0 }

\catcode`@=11
\newcount\r@fcount \r@fcount=0
\newcount\r@fcurr
\immediate\newwrite\reffile
\newif\ifr@ffile\r@ffilefalse
\def\w@rnwrite#1{\ifr@ffile\immediate\write\reffile{#1}\fi\message{#1}}

\def\writer@f#1>>{}
\def\referencefile{
  \r@ffiletrue\immediate\openout\reffile=\jobname.ref%
  \def\writer@f##1>>{\ifr@ffile\immediate\write\reffile%
    {\noexpand\refis{##1} = \csname r@fnum##1\endcsname = %
     \expandafter\expandafter\expandafter\strip@t\expandafter%
     \meaning\csname r@ftext\csname r@fnum##1\endcsname\endcsname}\fi}%
  \def\strip@t##1>>{}}

\def\citeall#1{\xdef#1##1{#1{\noexpand\cite{##1}}}}
\def\cite#1{\each@rg\citer@nge{#1}}     

\def\each@rg#1#2{{\let\thecsname=#1\expandafter\first@rg#2,\end,}}
\def\first@rg#1,{\thecsname{#1}\apply@rg}       
\def\apply@rg#1,{\ifx\end#1\let\next=\relax
\else,\thecsname{#1}\let\next=\apply@rg\fi\next}

\def\citer@nge#1{\citedor@nge#1-\end-}  
\def\citer@ngeat#1\end-{#1}
\def\citedor@nge#1-#2-{\ifx\end#2\r@featspace#1 
  \else\citel@@p{#1}{#2}\citer@ngeat\fi}        
\def\citel@@p#1#2{\ifnum#1>#2{\errmessage{Reference range #1-#2\space is bad.}
    \errhelp{If you cite a series of references by the notation M-N, then M and
    N must be integers, and N must be greater than or equal to M.}}\else%
 {\count0=#1\count1=#2\advance\count1
by1\relax\expandafter\r@fcite\the\count0,%
  \loop\advance\count0 by1\relax
    \ifnum\count0<\count1,\expandafter\r@fcite\the\count0,%
  \repeat}\fi}

\def\r@featspace#1#2 {\r@fcite#1#2,}    
\def\r@fcite#1,{\ifuncit@d{#1}          
    \expandafter\gdef\csname r@ftext\number\r@fcount\endcsname%
    {\message{Reference #1 to be supplied.}\writer@f#1>>#1 to be supplied.\par
     }\fi%
  \csname r@fnum#1\endcsname}

\def\ifuncit@d#1{\expandafter\ifx\csname r@fnum#1\endcsname\relax%
\global\advance\r@fcount by1%
\expandafter\xdef\csname r@fnum#1\endcsname{\number\r@fcount}}

\let\r@fis=\refis                       
\def\refis#1#2#3\par{\ifuncit@d{#1}
    \w@rnwrite{Reference #1=\number\r@fcount\space is not cited up to now.}\fi%
  \expandafter\gdef\csname r@ftext\csname r@fnum#1\endcsname\endcsname%
  {\writer@f#1>>#2#3\par}}

\def\r@ferr{\endreferences\errmessage{I was expecting to see
\noexpand\endreferences before now;  I have inserted it here.}}
\let\r@ferences=\references
\def\references{\r@ferences\def\endmode{\r@ferr\par\endgroup}}

\let\endr@ferences=\endreferences
\def\endreferences{\r@fcurr=0
  {\loop\ifnum\r@fcurr<\r@fcount
    \advance\r@fcurr by 1\relax\expandafter\r@fis\expandafter{\number\r@fcurr}%
    \csname r@ftext\number\r@fcurr\endcsname%
  \repeat}\gdef\r@ferr{}\endr@ferences}


\let\r@fend=\endpaper\gdef\endpaper{\ifr@ffile
\immediate\write16{Cross References written on []\jobname.REF.}\fi\r@fend}

\catcode`@=12

\citeall\refto          
\citeall\ref            %
\citeall\Ref            %

\vglue 1. truein
\nopagenumbers

\tutp{94-11}

\

{\noindent \bf
ELECTROWEAK STRINGS, SPHALERONS AND MAGNETIC FIELDS
}

\smallskip

\hskip 0.5 truein {Tanmay Vachaspati}

\vskip 0.1 truein

{
\obeylines \singlespace
\hskip 0.5 truein Tufts Institute of Cosmology
\hskip 0.5 truein Department of Physics and Astronomy
\hskip 0.5 truein Tufts University, Medford, MA 02155.
}

\vskip 0.4 truein

\body



\oneandathirdspace

\noindent{\bf 1. INTRODUCTION}

In this talk I would like to discuss three main topics. First, I will
connect electroweak strings and the sphaleron in a more
elaborate manner than that described in my paper with George
Field\refto{tvgf} and discuss the important issue of the stability
of electroweak strings.
Secondly, I would like to initiate a discussion of the
formation of electroweak strings and their evolution by painting a heuristic
picture of the phase transition if the  phase transition is weakly first
order or second order. Inextricably tied to this picture is a possible scenario
for baryogenesis and it is hoped that the present discussion will inspire
more rigorous work to either confirm or reject the scenario. Finally, I will
make a few comments on the generation of magnetic fields at the
electroweak phase transition. Kari Enqvist has already described the basic
idea of how this scenario works\refto{ke} and the purpose of these comments
is to clarify some of the issues that have not been addressed in
the literature. I will also describe the scenario in an alternate way that
makes an interesting connection between electroweak strings and
primordial magnetic fields.

\noindent{\bf 2. THE Z-STRING AND THE SPHALERON}

The bosonic sector of the standard model of the electroweak interactions is
described by the Lagrangian:
$$
L = L_W + L_B + L_{\Phi} - V(\Phi )
\eqno (2.1)
$$
where,
$$
L_W = - {1 \over 4} W_{\mu \nu a} W^{\mu \nu a}
\eqno (2.2)
$$
$$
L_Y = - {1 \over 4} Y_{ \mu \nu} Y^{\mu \nu}
\eqno (2.3)
$$
where $W_{\mu \nu}^a$ and $Y_{\mu \nu}$ are the field
strengths for the $SU(2)$ and $U(1)$ gauge fields $W_\mu ^a$ and $Y_\mu$
respectively. Also,
$$
L_\Phi = |D_\lambda \Phi |^2 \equiv
   \left |\left (\partial _\lambda - \half ig \tau ^a W_\lambda ^a - \half
                i g' Y_\lambda \right ) \Phi \right | ^2
\eqno (2.4)
$$
$$
V(\Phi ) = \lambda (\Phi ^{\dag} \Phi - \eta ^2 /2 )^2 \ ,
\eqno (2.5)
$$
where, $\Phi$ is a complex doublet.
In addition, we define
$$
\vec Z \equiv cos\theta_W n^a {\vec W} ^a - sin\theta_W \vec Y \ ,
\ \ \ \
\vec A \equiv sin\theta_W n^a {\vec W} ^a + cos\theta_W \vec Y \ ,
\eqno (2.6)
$$
where,
$n^a$ is the unit vector
$$
n^a = - {{ \Phi^{\dag} \tau^a \Phi} \over {\Phi^{\dag} \Phi}}
\eqno (2.7)
$$
$$
tan\theta_W \equiv {{g'} \over g} \ , \ \ \ \
\alpha \equiv (g^2 + {g'} ^2 )^{1/2} \ .
\eqno (2.8)
$$

The electroweak model has two different string solutions\refto{tvmb, mbtvmb},
namely, the $W-$string and the $Z-$string\refto{nambu, nm, tv1}. The latter
is the lighter of the two and
is better studied and hence we will not discuss the $W-$string but only
comment that some of the ensuing discussion directly applies
to the $W-$string as well.

The $Z-$string is a solution to the classical equations of motion following
from the Lagrangian in (2.1). In cylindrical coordinates $(r, \theta , z)$ it
is:
$$
\Phi = {{\eta} \over {\sqrt{2}}} f(r) e^{i m \theta  } \pmatrix{0\cr 1\cr} \ ,
\ \ \ \
Z_\mu = - {{v(r)} \over {\alpha r}} \delta_{\mu \theta}
\eqno (2.9)
$$
with all the other gauge fields equal to zero. The functions $f(r)$ and
$v(r)$ can be evaluated numerically (for example, see Ref. \cite{plrm}).
The string solution in (2.9) is precisely a Nielsen-Olesen vortex\refto{hnpo}
that has been embedded in the electroweak model\refto{tvmb, mbtvmb}.


The field configuration in (2.9) describes a tube of energy density
localized around the z-axis. At the center of this tube the Higgs field
vanishes. The tube contains a flux of Z-magnetic field whose value is:
$$
F_Z = {{4 \pi } \over {\alpha}} =
          {{4\pi} \over e} sin\theta_W ~cos\theta_W \ .
\eqno (2.10)
$$
where, $e = g sin\theta_W$ is the electric charge on an electron.
The energy per unit length of the string is:
$$
\mu = \pi \eta ^2 M(\beta )
\eqno (2.11)
$$
where $M(1) = 1$ and the tendency of $M$ is to increase with increasing
$\beta$. But the dependence of $M$ on $\beta$ is not too strong in the
neighbourhood of $\beta = 1$ and so we shall set $M=1$ whenever
we need numerical estimates. (For a plot of $\mu$ as a function of $\beta$,
see Ref. \cite{plrm}.)

The solution in (2.9) describes an infinite $Z-$string along the z-axis
but one can think of finite string configurations in the form of
closed loops\refto{marty} or open segments. Finite string configurations
in vacuum would not be static but would oscillate. And if the configuration
is large compared to the string thickness, the dynamics is described quite
accurately by the Nambu-Goto action. With time such oscillating loops and
segments will radiate away their energy and convert into particles. If the
loops and segments are not in vacuum but in a plasma that strongly interacts
with the strings, the dynamics will be very different and one expects that
the oscillations will be damped and the radiation suppressed.

A finite segment of $Z-$string terminates on magnetic monopoles\refto{nambu}.
The electromagnetic flux emanating from a monopole is:
$$
F_A = {{4\pi} \over {\alpha}} tan\theta_W
                   = {{4\pi} \over e} sin^2 \theta_W \ .
\eqno (2.12)
$$
One way to understand the presence of monopoles at the end of $Z-$strings
is to note that the $Z$ gauge field is a linear superposition of the
$W^3$ and $Y$ fields as given in eqn. (2.6). Then, when the string terminates,
the $Y$ flux cannot terminate because it is a $U(1)$ gauge field and the
$Y$ magnetic field is divergenceless. Therefore some field must continue
even beyond the end of the string. This has to be the massless field
of the theory, that is, the electromagnetic field.

Following Nambu, the asymptotic Higgs field configurations of a
monopole and an antimonopole are\refto{nambu}:
$$
\Phi_m = \pmatrix{ cos(\theta_m / 2) \cr
                     sin(\theta_m /2) ~e^{i\phi }\cr } \ , \ \ \ \
\Phi_{\bar m} = \pmatrix{ sin(\theta_{\bar m} /2) \cr
                          cos(\theta_{\bar m} /2) ~ e^{i\phi}}
\eqno (2.13)
$$
where $\theta_m$ and $\theta_{\bar m}$ are spherical angles defined with the
monopole and antimonopole at the origin respectively and we
have rescaled the Higgs field so that the vacuum manifold is given by
$\Phi^{\dag} \Phi = 1$. The
gauge fields are taken to be so that the covariant derivative of the Higgs
field vanishes.
$$
g W_\mu ^a = - \epsilon^{abc} n^b \partial_\mu n^c + i cos^2 \theta_W
 n^a (\Phi^{\dag} ~ \partial_\mu \Phi - \partial_\mu \Phi^{\dag} ~\Phi )
\eqno (2.14a)
$$
$$
g' Y_\mu = - i sin^2 \theta_W (\Phi^{\dag} ~ \partial_\mu \Phi -
                                    \partial_\mu \Phi^{\dag} ~\Phi ) \ .
\eqno (2.14b)
$$
where, $n^a$ is defined in (2.7).
For the special case $\theta_W = 0$, this is equivalent to the usual
$$
W_\mu = - i {{2} \over g} ( \partial_\mu U ) U^{-1}
\eqno (2.15)
$$
where, $U$ is a $2\times 2$ unitary matrix defined by
$$
\Phi = U \pmatrix{1, 0}^T
\eqno  (2.16)
$$
Note that the monopole configuration has a singularity along the negative
z-axis since the Higgs field becomes multi-valued when we set
$\theta_m = \pi$. Similarly the antimonopole has a singularity along the
positive z-axis ($\theta_{\bar m} = 0$). These singularities tell us the
location of the
$Z-$string that is attached to the monopole and the antimonopole.

Once we have the monopole and antimonopole configurations, we can patch them
together to get the field configuration for a finite segment of $Z-$string:
$$
\Phi_{m\bar m} = \pmatrix{cos(\Theta /2) \cr
                          sin(\Theta /2) ~ e^{i\phi} \cr }
\eqno (2.17)
$$
where,
$$
cos\Theta \equiv cos\theta_m - cos\theta_{\bar m} +1 \ .
\eqno (2.18)
$$
It is straightforward to check that (2.17) yields the monopole field
configuration close to the monopole ($\theta_{\bar m} \rightarrow 0$)
and the antimonopole configuration close to the antimonopole
($\theta_{m} \rightarrow \pi$). It also yields a string singularity
along the straight line joining the monopole and antimonopole
($\theta_m = \pi , ~ \theta_{\bar m} = 0$).

What is important for us is that there are other Higgs field
configurations that also describe monopoles and antimonopoles. These are
given by global $U(1)$ transformations of (2.13). Therefore we will write
$$
\Phi_m = e^{i\gamma} \pmatrix{ cos(\theta_m /2) \cr
                     sin(\theta_m /2) ~e^{i\phi }\cr } \ , \ \ \ \
\Phi_{\bar m} = e^{i\gamma} \pmatrix{ sin(\theta_{\bar m} /2) \cr
                          cos(\theta_{\bar m} /2) ~ e^{i\phi}}
\eqno (2.19)
$$
This seemingly trivial observation is very useful because it allows us
to construct $Z-$string segments which are twisted. Consider the Higgs
field configuration:
$$
\Phi_{m \bar m} (\gamma ) =
\pmatrix{
            sin(\theta_m /2) sin(\theta_{\bar m} /2) e^{i\gamma}
          + cos(\theta_m /2) cos(\theta_{\bar m} /2) \cr
   sin(\theta_m /2) cos(\theta_{\bar m} /2) e^{i\phi }
 - cos(\theta_m /2) sin(\theta_{\bar m} /2) e^{i(\phi - \gamma )}
                                                               \cr
}
\eqno (2.20)
$$
together with the gauge fields given by eqn. (2.14). When we take the limit
$\theta_{\bar m} \rightarrow 0$ we find the monopole configuration of
(2.13) and when we take $\theta_m \rightarrow \pi$ the configuration is
that of the antimonopole of eqn. (2.19) provided we perform the rotation
$\phi \rightarrow \phi + \gamma$. The monopole and antimonopole in (2.20)
also have the usual string singularity joining them. This means that
the configuration in (2.20) describes a monopole and antimonopole pair
that are joined by a $Z-$string segment that is twisted by an angle
$\gamma$.

Now we will calculate the Chern-Simons number (which is, loosely speaking,
the baryon number) of the twisted segment of string described in (2.20)
and with gauge fields in (2.14).

Let me assume that $\gamma$ is a rational fraction of $2\pi$ so that
we can write $\gamma = 2\pi p/q$ where $p$ and $q$ are integers. Then we
can take $q$ twisted segments, each of which is described by eqns. (2.20) and
(2.14), and join them up - the antimonopole of one segment
can be brought to annihilate the monopole of another segment - to form
a closed loop. In this way we will get a loop of $Z-$string that is twisted
by an angle $2\pi p$. Now we need to calculate the Chern-Simons number of
this loop. The Chern-Simons number is defined as
$$
CS =
{{N_F} \over {32\pi^2}}  \int  d^3 x \epsilon_{ijk}
 \biggl [ g^2 \biggl ( W^{a ij} W^{a k} - {g \over 3}
                       \epsilon_{abc} W^{ai} W^{bj} W^{ck} \biggr ) -
          {g'}{}^2 Y^{ij} Y^{k} \biggr ] \ .
\eqno (2.21)
$$
where, we have included the number of families $N_F$ in the definition.
For a closed loop of $Z-$string, this expression simplifies considerably
since the only non-vanishing gauge field
is the $Z$ gauge field. In this circumstance, (2.21) reduces to
$$
CS = N_F {{\alpha^2} \over {32\pi^2}} cos(2\theta_W )
      \int d^3 x ~{\vec Z} \cdot ( {\vec \nabla} \times {\vec Z} )
\eqno (2.22)
$$

We can now easily calculate the Chern-Simons number
of the twisted loop by using the result that if we have a twisted flux loop
of a gauge field $\vec A$ with flux $F$ and twist $2\pi p$, then,
$$
\int d^3 x ~{\vec A} \cdot ( {\vec \nabla} \times {\vec A}) ~ = ~ 2 ~F^2 ~p \ .
\eqno (2.23)
$$
This result is well-known to people working in hydrodynamics and
astrophysics\refto{mbgf} but less known in the particle physics
community. So we give a quick sketch of the derivation.

We first note that the internal twisting of a flux tube is
equivalent to a linking of two different flux tubes\refto{knots}. Hence
we can restrict ourselves to evaluating the left-hand side of (2.23) for
two untwisted flux tubes that are linked $p$ times.

We work in the gauge ${\vec \nabla} \cdot {\vec A} = 0$ and the
magnetic field is given by ${\vec B} = {\vec \nabla} \times {\vec A}$.
Therefore, if we are given a magnetic field configuration, the gauge
field can be found from
$$
\vec A (\vec x ) = - {1 \over {4\pi}} \int d^3 x' {\vec B} ( {\vec x}{}' )
 \times {{{\vec x} - {{\vec x}{}'}} \over {| {\vec x} - {{\vec x}{}'}|}} \ .
\eqno (2.24)
$$
Denoting the integral on the left-hand side of (2.23) by $I$, we then have:
$$
I = - {1 \over {4\pi}}
\int d^3 x ~{\vec B} (\vec x ) \cdot \int d^3 x' {\vec B} ( {\vec x}{}' )
  \times {{{\vec x} - {{\vec x}{}'}} \over {| {\vec x} - {{\vec x}{}'}|}}
\eqno (2.25)
$$
and then perform
integrations over the cross-section of the flux tubes. This has the effect
$$
I = (2 F^2)
\left [ - {1 \over {4\pi}}
\oint d{\vec f} \cdot \oint d{\vec g}
  \times {{{\vec f} - {\vec g}} \over {| {\vec f} - {\vec g} |} }
\right ]
\eqno (2.26)
$$
where, we have assumed that both flux tubes carry the same flux $F$ and
that their locations are given by $\vec f$ and $\vec g$.
The expression in square brackets is the Gauss linkage formula\refto{flanders}
for the curves $\vec f$ and $\vec g$ and this proves (2.23).

Now, using (2.23) (with $\vec A$ replaced by $\vec Z$) in (2.22) and
inserting the value of the $Z$ flux in the string (eqn. (2.10)), we get the
Chern-Simons number of the $Z-$string loop that is twisted by $2\pi p$:
$$
CS = {N_F} cos(2\theta_W ) p
\eqno (2.27)
$$
Since the loop was built out of $q$ segments and $\gamma = 2\pi p/q$, the
Chern-Simons number of one segment is
$$
CS = N_F cos2\theta_W ~ {{\gamma} \over {2\pi}} \ .
\eqno (2.28)
$$

So far we have been working with string segments having arbitrary twist.
But now consider the case, $\gamma = \pi /cos(2\theta_W )$. With this
twist, the Chern-Simons number is $N_F /2$ - precisely that of the
sphaleron\refto{minos}!

Given that the segment with twist $\pi /cos(2\theta_W )$ has
Chern-Simons number equal to that
of the sphaleron, it is natural to ask if some deformation
of it will yield the sphaleron. This deformation is not hard to guess for
the $\theta_W = 0$ case. In this case, if we let the segment size shrink
to zero, we have $\theta_m = \theta_{\bar m} = \theta$ and the Higgs field
configuration of (2.20) gives:
$$
\Phi_{m \bar m} (\gamma = \pi ) =
               \pmatrix{cos\theta\cr sin\theta ~ e^{i\phi}\cr} \ .
\eqno (2.29)
$$
And this is exactly the field configuration of Nick Manton's $SU(2)$
sphaleron\refto{nm}. (Note that the gauge fields continue to be given
by (2.14) or, equivalently in this case, by (2.15).)

Encouraged by this successful connection between the twisted string segment
and the sphaleron in the $\theta_W = 0$ case, we can conjecture that the
twisted segment of $Z-$string with Chern-Simons number $N_F /2$ will
collapse into the sphaleron {\it for any $\theta_W$}. Therefore the asymptotic
Higgs field configuration for the sphaleron can be obtained by letting
$\theta_m = \theta_{\bar m} = \theta$ in (2.20) and taking
$\gamma = \pi /cos(2\theta_W ) = \gamma_S$.
Denoting the sphaleron asymptotic Higgs field
configuration by $\Phi_S$, we conjecture
$$
\Phi_S = \pmatrix{
 sin^2(\theta /2) - cos^2(\theta /2) ~ e^{i\gamma_S}\cr
 sin(\theta /2) cos(\theta /2) ~ e^{i\phi} (1 - e^{-i\gamma_S})\cr
                  }
\eqno (2.30)
$$
and the asymptotic gauge fields are given by (2.14).
If this conjecture is true, we should
be able to find a solution to the classical equations of motion with this
ansatz for the asymptotic fields. Of course, we would first need to insert a
suitable radial dependence in the ansatz and then solve the field equations.

On physical grounds it seems reasonable that there should be a critical
value of twist at which one can get a static solution for a $Z-$string
segment. This is because the segment likes to shrink under its own tension
but the twist prevents the shrinkage and is equivalent to a repulsive
force between the monopole and antimonopole. Then, if the string is
sufficiently twisted, the attractive force due to the tension and
the repulsive force due to the twist will balance and a static solution
can exist. So far we have been assuming
that the only dynamics of the segment is towards collapsing or expanding.
However, since we are dealing with twisted segments, we should also
include the rotational dynamics associated with twisting and untwisting.
So, while any twist greater than a certain critical twist will successfully
prevent the segment from collapsing, only a special value of the twist
will give a static solution to the rotational dynamics. Furthermore,
we expect that this solution will be unstable towards rotations that
twist and untwist the string segment. This would be the unstable mode of
the sphaleron.

A question that I have frequently been asked is that how can an object have
fractional Chern-Simons number? This question is easily answered: only the
vacuua have integer Chern-Simons number; outside the vacuua, the Chern-Simons
number can be anything. But then, what would happen when this object
having fractional Chern-Simons number decays into particles? We know that
the change in the Chern-Simons number of the object equals the change in the
baryon number, then how can fractional baryon number be produced?
The answer to this question is currently under investigation by Eddie Farhi
and collaborators\refto{efarhi} and requires a quantum treatment of
the decay of the
fractionally charged object. The belief is that the equality of baryon number
change to the change in Chern-Simons number should be viewed as an equality
of expectation values. So the change in baryon number will always be
integral and only upon averaging will one get a fractional change in baryon
number.

This completes what I wanted to say about the connection of the $Z-$string
and the sphaleron. (Other discussions of this connection may be found in
Refs. \cite{mbtvmb, mhmj}.)
Just to summarize, twisted segments of $Z-$string carry
Chern-Simons number in proportion to their twist and when the twist is
$\pi / cos(2\theta_W)$, the Chern-Simons number is $N_F /2$. We conjecture
that the string segment with this special value of twist is precisely an
extended sphaleron.

\noindent{\bf 3. $Z-$STRING STABILITY}

A non-trivial and very important aspect of studying electroweak strings is to
understand their (meta-)stability. In an exhaustive analysis done in
collaboration with Margaret James and Leandros Perivolaropoulos, we plotted
the region of parameter space in which the $Z-$string is stable towards small
perturbations\refto{mjlptv}.
The conclusion is that one can only have stable strings for
$sin^2 \theta_W \gtwid 0.9$, that is, for $g' \gtwid 3 g$. The experimentally
determined value of $sin^2 \theta_W$ is $0.23$ and is deep inside the
instability region.

Recently there have been a few papers that have shed more light on the
issue of stability for $\theta_W$ not too large. First there was a paper by
Warren Perkins that showed that there is an instability towards developing
$W$ fields\refto{wp} for $sin^2 \theta_W \ltwid 0.8$. Then
Manuel Barriola, Martin Bucher and I used a simple argument
to show that the string is unstable in the case $\theta_W = \pi /4$.
In this instability, the upper component of the Higgs field grows,
the lower component diminishes in magnitude and the gauge field
spreads\refto{mbtvmb}. To see the explicit form of the instability,
denote the unperturbed electroweak string solution by $\Phi_0$ and
$Z_j ^{(0)}$ and consider the sequence of field configurations labeled
by a parameter $\xi$:
$$
\eqalign{
\Phi (\vec x ;\xi )&=
\cos \xi \ \Phi _0(\cos \xi \ \vec x )+\sin \xi \ \Phi _\bot \cr
Z_j (\vec x ;\xi )&=\cos \xi \ Z_j ^{(0)} (\cos \xi \ \vec x )\cr }
\eqno (3.1)
$$
where, $Z_j$ is given in terms of the $SU(2)_L \times U(1)_Y$ gauge fields
in eqn. (2.6) together with $n^a = (0,0,1)^T$ (note!), and,
$$
\Phi _\bot = {{\eta} \over {\sqrt{2}}} \pmatrix{1\cr 0\cr} \  .
\eqno (3.2)
$$
For $\xi = 0$, the configuration is the embedded defect solution
and for $\xi = \pi /2$, the configuration describes the vacuum.
We then have,
$$
\eqalign{
\Phi^ {\dag}(\vec x ;\xi ) \Phi (\vec x;\xi )&
= \cos ^2 \xi \ {\Phi_0}^{\dag} \Phi_0 + \sin ^2 \xi ~ \eta ^2
\cr
W_{ij}^a (\vec x ;\xi ) W_{ij}^a (\vec x ;\xi )
&= \cos ^4 \xi \  W_{ij}^a(\vec x ) W_{ij}^a(\vec x ),
\cr
Y_{ij}^a (\vec x ;\xi ) Y_{ij}^a (\vec x ;\xi )
&= \cos ^4 \xi \  Y_{ij}^a(\vec x ) Y_{ij}^a(\vec x ),
\cr
D_i \Phi  (\vec x ;\xi )= \cos ^2 \xi &\left [ \partial _i +
i{\alpha \over 2} Z_i ( \vec x ){\cal T}^{Z} \right ] \Phi_0(\vec x)
\cr
V[\Phi (\vec x , \xi )] &= \cos ^4 \xi \  V[\Phi _0(\vec x )]
\cr
}
\eqno (3.3)
$$
where,
$$
{\cal T}^{Z} = \pmatrix{0&0\cr 0&1\cr}
\eqno (3.4)
$$
{}From (3.3) we get
$$
[D_i \Phi ( \vec x ;\xi )]^\dagger  \ [D_i \Phi (\vec x ;\xi )]
=\cos ^4\xi \ (D_i \Phi _0 (\vec x ))^\dagger  (D_i \Phi _0 (\vec x )) \ ,
\eqno (3.5)
$$
and hence, the total energy of the configuration is
$$
E(\xi )=\cos ^{2}\xi \ E(\xi =0) \ .
\eqno (3.6)
$$
This explicitly shows a sequence of configurations that start at
the string ($\xi = 0$) and end up in the vacuum ($\xi = \pi /2$)
and with monotonically decreasing energy.

Although in the sequence of configurations considered above, only
the $Z$ field is written as being non-zero, this is really misleading because
$\Phi$ is no longer proportional to $(0, 1)^T$. So the gauge
fields in the above configurations actually include $W^{\pm}$ fields
and the instability demonstrated above is towards developing
$W$ fields. If the parameter $\xi$ is replaced by a parameter
along the length of the string ($z$), the above sequence of configurations
are slices of a string along the $z$ axis that has terminated in a monopole.
So the instability described above is the instability of a string to
break-up by forming monopoles and antimonopoles.

Frans Klinkhamer and Poul Olesen\refto{fkpo}, and, independently, Margaret
James\refto{james1}, have
constructed a two parameter family of configurations - a two sphere
in configuration space - with the string on top of the sphere for
$sin^2 \theta_W \ltwid 0.7$. This picture implies that the string has at
least two unstable modes for small $\theta_w$.
However, it seems likely that there is only one unstable mode of the
string solution under {\it small} perturbations since the two modes
that Klinkhamer and Olesen, and, James, have found are gauge equivalent
when the perturbations are taken to be small at the solution\refto{james2}.
If this is true, we might reasonably guess that the only unstable
perturbative mode is towards breaking up of strings by forming
monopole-antimonopole pairs\footnote*{The physics of the two unstable
modes is not completely clear to me. Perhaps the constructions of
Ref. \cite{fkpo, james1} are counting the instabilities towards
forming monopoles that are twisted in the way described in Sec. 2.}.

There is a simple way to understand why the $W$ fields develop\refto{japo}
and destabilize the $Z$ string\refto{poprivate, wp}.
For this consider the energy of a spin
$s$ particle in a uniform magnetic field $\vec B$ along the z-direction:
$$
E^2 = k_z^2 + m^2 + (2n+1) e B - 2 e {\vec B}\cdot {\vec s}
\eqno (3.7)
$$
where, the charge on the particle is $e$ and the g-factor of the particle
(in the last term) is taken to be 2.
The right-hand side contains the z-momentum and mass contribution to
the energy. Then there is the contribution of the various Landau levels
labeled by the integer $n$ and finally there is the coupling of the
magnetic moment (spin) to the magnetic field.
Clearly if $s=1$, it is possible for the right-hand side to be negative
if
$$
B > {m^2 \over e}
\eqno (3.8)
$$
and hence there is an instability of the magnetic field to condensing
spin one particles or, to ``$W$ condensation''.

In the case of the $Z-$string, the magnetic field $\vec B$ in (3.7)
is the $Z-$magnetic field within the string and the spin-1 particles are
the $W$ bosons. It can be checked that the g-factor is indeed 2 as has
been assumed in writing (3.7) and the $Z$ charge on the $W$ boson is
$e_Z = g cos\theta_W$. Therefore the conditions are right for the
$W$ condensation
argument to apply to the $Z$ string. The only condition that is not
immediately satisfied is the assumption in (3.7) that the magnetic field is
uniform. But if the string is not too thin relative to the $W$ boson Compton
wavelength, one is justified in thinking of the $Z$ magnetic field of the
string as being uniform. This is precisely the case for low values of
$sin^2 \theta_W$. For small values of $\theta_W$, we also see that
$e_Z = g cos\theta_W $ is not small and so the critical $Z$ magnetic
field required for $W$ condensation (eqn. (3.8)) can be relatively
small. These facts show that $Z$ strings are unstable
towards $W$ condensation for small $\theta_W$.

A point to remember is that we have only been considering infinite
strings so far. The instability towards break-up will not apply
when the string length is less than a certain critical length since
the formation of monopoles requires certain  gradients along the
length of the string. The critical length of string below which
break-up is not possible is likely to be a few times the size of the
monopole ($\sim m_W^{-1}$). Such short segments are not string-like
in appearence but can still be important as they can carry Chern-Simons
number just as in the case of longer segments.

Are there any new instabilities of a finite segment that are absent
in the case of the infinite string? As the string segment is not topological,
there is a possibility that the string segment could destabilize as
a whole and disappear into the vacuum. But this does not appear to be
possible since the evolution of the string segment has to conserve
magnetic charge and so the
only way a monopole can disappear is to annihilate with an antimonopole.
Therefore the only new instability of a finite segment of string
is towards dynamical collapse. If the segment has angular momentum
or is twisted, this dynamical instability may not be too severe.
(Nambu has given rough estimates for the life-time of a long
string segment with angular momentum towards radiative decay\refto{nambu}.)

There are some circumstances under which the stability of the $Z-$string
improves. Thermal effects can be shown to improve stability\refto{holmanetal}
but the improvement is more so that strings with higher values of
$\beta$ become stable and strings with small values of $\theta_W$ remain
unstable. Rick Watkins and I showed that if there are particles
bound to the string, the string can be stable down to lower values of
$sin^2\theta_W$. In our analysis\refto{tvrw}, we found
stable strings down to
about $sin^2\theta_W = 0.5$ but did not find stable $Z-$strings
with $sin^2\theta_W = 0.23$.

The picture then is the following: if we start with a long (infinite)
$Z-$string in the case when $\theta_w$ is small, the string will rapidly
break up into small segments of a critical length - of the order of
a few times the string thickness - which will then survive
until they radiate away their energy and angular momentum.

If we are willing to consider extensions of the standard model, stable
strings are possible. The popular two Higgs model
does not yield stable strings\refto{memj} but more complicated models,
such as, left-right models can yield stable
strings\refto{holmanetal}. Other extensions can also give topologically
stable strings with $Z-$flux in them\refto{gdgs}.

\noindent{\bf 4. ELECTROWEAK STRINGS AT THE PHASE TRANSITION}

I will assume that the electroweak phase transition is second order
or weakly first order - after hearing the various talks at this
conference, this seems likely.
This means that the correlation length at the phase
transition is of the order $T_c^{-1}$ where $T_c$ is the critical
temperature at which the phase transition occurs. Under this assumption
the temperature ($T$) dependent vacuum expectation value of $\Phi$
is given by:
$$
<| \Phi |> = {\eta \over \sqrt{2}}
             \left ( 1 - {{T^2} \over {T_c^2}} \right )^{1/2} \ .
\eqno (4.1)
$$
Therefore the mass per unit length of the $Z-$string follows from
(2.11):
$$
\mu (T) \approx \pi M(\beta ) <| \Phi |>^2 = {{\pi \eta^2} \over {2}}
             \left ( 1 - {{T^2} \over {T_c^2}} \right ) M(\beta )\ .
\eqno (4.2)
$$
Let us now define the ``Hagedorn temperature'' $T_H$ as the temperature
at which
$$
T_H = \sqrt{{3 \mu (T_H)} \over {2\pi}} \ .
\eqno (4.3)
$$
This gives,
$$
T_H = T_c \sqrt{{x \over {1+x}}} \ , \ \ \ \ {\rm with} ~ ~ ~
x \equiv {{3M(\beta ) \eta^2} \over {4 T_c ^2}} \ .
\eqno (4.4)
$$
For $\beta = 1$ (that is, $m_H = m_Z$), $\eta = 250 GeV$ and $T_c = 100 GeV$,
this gives $T_H \approx 0.9 T_c = 90 GeV$. Note that, while $T_H$ is quite
close to $T_c$, the time required for the universe to cool from $T_c$ to
$T_H$ is about $10^{16} T_H^{-1}$.

The Hagedorn temperature is significant in the study of string statistical
mechanics because it is the temperature that marks a phase transition in
a string network. Below the Hagedorn temperature, a string network
prefers to break up into the smallest possible loops or segments. Above
the Hagedorn temperature, the network is dominated by long strings.
These results have been obtained by studying the statistical mechanics
of a box of strings and it is believed that the results apply to fundamental
strings as well as strings that arise as defects in a phase transition.

How can we understand these results? The basic point is that the number
of states of a string grows exponentially with length while the Boltzmann
distribution decays exponentially with length. It is found that the partition
function depends on an integrand proportional to
$$
exp\left [ \left ( \sqrt{{2\pi} \over {3 \mu (T)}}   - {1 \over T}
    \right ) E \right ]
\eqno (4.5)
$$
where $E$ is the energy of the string. Above the Hagedorn temperature,
the number of states associated with the long strings wins over the
Boltzmann suppression and the network settles into long (infinite) strings.

There are various subtleties in studying string statistical mechanics
and the above argument is only meant to suggest that something strange
should happen above the Hagedorn temperature which is still below the
critical temperature.  For those wishing to pursue this line of argument
further, a paper summarizing earlier
work on the statistical analysis of strings as well as applying string
statistical mechanics to cosmic strings, is the paper by Dave Mitchell and
Neil Turok\refto{dmnt}.  In this paper, the authors also treat the
statistical mechanics of open strings where the monopoles at the end of
strings and the strings themselves form during the same cosmological phase
transiton - exactly the case relevant to electroweak strings! However,
Mitchell and Turok ignore string-string interactions in their anlaysis and
this may be a bad approximation to make.
In particular, important processes such as
string break-up and reconnections have not been taken into account in the
analysis. For this reason, we cannot import the results of Ref. \cite{dmnt}
wholesale but can only take them as an indication that the density of
strings may be significant at temperatures near the Hagedorn temperature
but still below the critical temperature.

There is another way of seeing that something bizzare may happen
above the Hagedorn temperature. For $T_c > T > T_H$, the tension in the string
is smaller than the thermal excitations due to interaction with the plasma.
Therefore the strings are ``hot'' and the tension in the string can be
ignored compared to thermal excitations.
This would mean that points on a string segment would diffuse and the
distance between any two points on a string with grow with time. So the
length of the segment would grow and a given segment would not
be able to shrink and disappear. All that a hot segment of electroweak
string can do is to grow in length and break up. The smaller pieces that
are formed due to the fragmentation would also not be able to shrink
and would keep growing bigger. Then the density of strings would become
high and the likelihood of strings merging to form bigger strings would
increase. In this way one is led to conclude that, at temperatures
lower than $T_c$ but higher than $T_H$, the strings will be relatively long.
This is in contrast to the picture where long strings are Boltzmann suppressed.

These arguments lead us to the picture that, for $T_c > T > T_H$, the plasma
contains a significant density of long strings. As the temperature
falls, the separation between strings increases.
If the phase transition is first order, the strings get
separated by the process of bubble formation whereas if the phase transition
is second order, the inter-string separation grows larger continuously.
At $T=T_H$ there is a phase transition in the string network itself and
long strings begin to break-up at this temperature. If the strings are
metastable,
the strings break-up by nucleating monopoles and antimonopoles and this
would be a first order phase transition in the string network itself. If
the strings are unstable, the break-up is more like a second order phase
transition in the string network. In this picture, the long electroweak
strings at $T_H$ are genuine relics of the unbroken phase and are not due
to thermal fluctuations (since these ceased at $T_c$).

Now we follow the evolution to temperatures below the Hagedorn
temperature. As the universe cools below the Hagedorn temperature, the
infinite strings
will break up and yield an exponential distribution of string segments
and the number of string segments of length $l$ will be
proportional to ${\rm exp}[- a l]$ where $a>0$ is a constant. With further
cooling, the tension
in the strings starts becoming important and the segments start shrinking
and decaying into radiation. All this will happen relatively quickly and
so we can say that the string network disappears at a temperature
$T \sim T_H$. As the network disappears, it will produce baryons and
anti-baryons since,
from our results in Sec. 2, twisted and linked strings carry baryon number.
If there is sufficient CP violation in the dynamics of the string
network, the decay of strings would lead to a production of net baryon
number.

The above scenario is a scenario of baryogenesis at the Hagedorn temperature
$T_H$ and it is a concern that subsequent sphaleron transitions might
erase any baryon number produced at $T_H$. To check if this happens,
we find the temperature $T_S$ at which the sphaleron
transition rate first falls below the Hubble expansion rate\refto{misha}:
$$
T_S {\rm exp}\left [{{- M_S (T_S)} \over T_S} \right ] =
               {{T_S ^2} \over {m_{Pl}} } \ .
\eqno (4.6)
$$
where, $M_S (T)$ is the temperature dependent mass of the
sphaleron\refto{fknm, kunz} and
$m_{Pl}$ is the Planck mass. Now we know that, for $\theta_W = 0$,
$$
M_S (T) = 2 B(\lambda / \alpha_W ) {{M_W (T)} \over {\alpha _W}}
\eqno (4.7)
$$
where, $B$ is a weakly dependent function of the coupling constant
ratio $\lambda / \alpha_W$, $\alpha_W = g ^2 /4\pi$
and $M_W (T)$ is the temperature dependent mass of the $W-$boson.
Writing $m_W = M_W(0)$ and approximating ${\rm ln}(T_S)$ by
${\rm ln}(T_c)$ we get,
$$
T_S = T_c \left [ 1 + \left \{
  {{\alpha_W T_c} \over {2 B m_W}} {\rm ln}
               \left ( {{m_{Pl}} \over T_c } \right ) \right \}^2
                     \right ] ^{-1/2} \ .
\eqno (4.8)
$$
Now comparing $T_S$ to $T_H$ (eqn. (4.4)), we find that $T_S$ is
larger than $T_H$ provided
$$
\left [ {{\alpha_W \eta} \over {4 m_W}} {\rm ln} \left (
                         {{m_{Pl}} \over {T_c}} \right ) \right ]^2
      3 M(\beta ) < [ B(\lambda /\alpha_W ) ]^2
\eqno (4.9)
$$
With $\alpha_W = 1/30$, $\eta = 250 GeV$, $m_W = 80 GeV$ and
${\rm ln}(m_{Pl} / T_c ) = 17 {\rm ln}(10)$, this condition gives
$$
      3 M(\beta ) < [ B(\lambda /\alpha_W ) ]^2 \ .
\eqno (4.10)
$$
The value of $B$ ranges from 1.5 to 2.7 - at least for the $SU(2)$
sphaleron (that is, for the $\theta_W = 0$ case) - and is roughly 1.9 when
$\beta = 1$ ($m_H = m_Z$) at which point $M(\beta ) =1$. Therefore
the condition (4.9) is satisfied in the parameter range of interest
and we have $T_H < T_S$. For $\theta_W \ne 0$ too, we expect that this
condition will be satisfied for an interesting range of parameters.

The above argument assures us that sphaleron transitions cannot
completely erase the
baryon number that would be generated by the string network at the Hagedorn
temperature. The exact fraction of baryon number that survives can
be found by studying the equations of detailed balance in an expanding
universe.

While the above scenario seems plausible to me, it has several
weaknesses. The first weakness is that we do not understand the formation
of electroweak strings at the electroweak phase transition. The arguments
given above suggest that electroweak strings should be present in
significant numbers above the Hagedorn temperature but we still do not
have a method to get a rigorous estimate of quantities
such as the number density of strings, or, the length distribution.
The second weakness is that
the production of baryons over antibaryons in the process of string
decay requires CP violation which is thought to be very weak in the
standard model. This, however, is a problem with any scenario of electroweak
baryogenesis and it is quite common to consider more strongly CP violating
extensions of the standard model. One could also take this approach
with electroweak strings. On the other hand, it would be more satisfying
if there were some unusual source of CP violation in the
monopole-antimonopole system within the standard model that may
not be present in the usual particle interactions. These issues are
presently being investigated.

\noindent{\bf 5. MAGNETIC FIELDS}

In this last section, I would like to discuss the generation of
cosmological magnetic fields arising at the electroweak phase
transition\refto{tvmag}.
Kari Enqvist has discussed the scenario in some detail and my purpose
is to make a few clarifying remarks about some of the assumptions that
go into the scenario. Then I would like to describe a connection between
electroweak strings and primordial magnetic fields.

The basic idea is that, as the Higgs field acquires a vacuum expectation
value, currents are produced that lead to a magnetic field. Specifically,
one defines the electromagnetic gauge field as in eqn. (2.6) but the
electromagnetic field strength as:
$$
\eqalign{
F_{\mu \nu} ^{em} = sin\theta _W n^a W_{\mu \nu} ^a +& cos\theta _W
            Y_{\mu \nu}
\cr
&
 - i4 g^{-1} \eta^{-2} sin\theta _W [ (D_\mu \Phi ) ^{\dag} D_\nu \Phi
                                 - (D_\nu \Phi ) ^{\dag} D_\mu \Phi ] \ .
\cr
}
\eqno (5.1)
$$

So far, all that we have done is to define the electromagnetic field
strength and have not included any dynamics. The dynamics comes in
when we estimate the various terms in (5.1) in a plasma at high
temperature. Suppose we are interested in a macroscopic volume of
size $L$ which is much larger than the thermal wavelength $T^{-1}$.
We expect that the field strengths $W_{\mu \nu}^a$ and $Y_{\mu \nu}$,
when averaged over the volume of size $L$,
will be exponentially decreasing with increasing $L$ since the plasma
is neutral on such scales and has no net currents either. However, the
covariant derivative is expected to fall off as a power law in $L$
as it is simply the covariant gradient of a scalar field. For example,
the covariant gradient may be estimated as the change in the value
of $\Phi$ occuring over the scale $L$ - therefore,
$|D_i \Phi | \sim \eta /L$.
This is the hinge on which the production of magnetic fields at the
electroweak phase transition rests and one can give several arguments
why this seems reasonable. (Though none of the arguments rigorously
proves the assumption.)

We now give the arguments that indicate that the covariant
derivative falls off as a power law and not exponentially with $L$.
The motivation for expecting
this behaviour comes from the Kibble argument. If the covariant
derivative vanishes, there is complete correlation between distant
parts of the universe which have never been in causal contact and
this is inadmissible. This argument is strictly valid for global
symmetries where the phases in $\Phi$ have physical meaning and
so the argument can legitimately
be questioned for gauge symmetries. On the other hand, note that
we are dealing with a system at high temperatures when the energy is
distributed among the various degrees of freedom. These include the
covariant derivative of $\Phi$ and it is natural to assume that the
covariant derivative takes on a value given by energy
equipartition. Whereas $W_{\mu \nu}^a$ and $Y_{\mu \nu}$ fall off
exponentially with $L$ due to vanishing currents in the plasma, there is no
current density in the particles (fermions etc.) in the plasma that
is responsible for the last term in (5.1) and so a power law fall off
is possible.

Here we will give another argument for why magnetic fields should be
produced at the electroweak phase transition.
This argument is consistent with the theme
of the talk as it suggests that there is a source for the magnetic
field and the source is none other than the electroweak monopoles present
at the ends of electroweak strings. As in Sec. 4, we expect
electroweak monopoles to be produced at the electroweak phase transition
and then to go away at a somewhat lower temperature. Therefore, imagine
a distribution of monopoles and antimonopoles in a plasma with the
magnetic lines of force running from monopoles to antimonopoles. In
addition, the monopoles are connected to antimonopoles via electroweak
strings. With time, the strings shrink and the monopoles and antimonopoles
at the ends of the string annihilate. However, the magnetic lines of
force are glued to the plasma because the plasma is a very good
electrical conductor. So the magnetic lines of force survive even when
the monopoles themselves annihilate and disappear. If the lines were
present only on very small scales, they too would eventually disappear
(as the conductivity of the plasma decreases with time). But this
is not likely; instead we
expect that the magnetic lines of force will percolate much like the
percolation of a cosmic string network\refto{tvavformation}. With
time, the small scale curvature on the magnetic lines of force will
disappear and the lines will straighten out but there will always
be lines of force present on large scales where they are frozen in
the plasma. Hence, at any epoch a relic magnetic field will be present.

The magnetic field strength is roughly given by $B \sim T^2$ where $T$ is the
temperature of the universe provided we assume that the field is frozen-in
on the smallest scales (of order $T^{-1}$).
However, this is an incorrect assumption
as the frozen-in scale is much larger than the thermal wavelength. If
the frozen-in scale is denoted by $l_f$, we have, $l_f \sim 10^{12} T^{-1}$.
Below this scale, the plasma is unimportant for the evolution of the magnetic
field and so the field will smooth itself out. Using a flux average to
estimate the field strength, we find $B \sim 10^{-12} T^2$ and this
field is coherent on a scale $l_f$. At the electroweak scale, this gives
$B \sim 10^{12} G$ with a coherence scale of $10^{-5} cms$.

The presence of strong magnetic fields in the very early universe leads
to very interesting physics and is currently under investigation by
several groups.
Here, I would only like to remark that, if the above
arguments connecting electroweak strings with magnetic fields is
correct, the direct or indirect observation of such a magnetic field
would immediately
yield information about the cosmological electroweak phase transition
and about electroweak strings!

\noindent {\bf Acknowledgements:}

I thank the organizers of this meeting and especially Filipe Freire for
hosting this conference and making it so enjoyable. This work was supported
by the National Science Foundation.


\references

\singlespace

\refis{holmanetal} R. Holman, S. Hsu,
T. Vachaspati and R. Watkins, {\it Phys. Rev.} {\bf D46}, 5352 (1992).

\refis{dmnt}
D. Mitchell and N. Turok, Nucl. Phys. B 294, 1138 (1987).

\refis{tvrw} T. Vachaspati and R. Watkins, Phys. Lett. B 318, 163 (1993).

\refis{tvgf} T. Vachaspati and G. B. Field, TUTP-94-1.

\refis{ke} K. Enqvist, at this conference.

\refis{tvmag} T. Vachaspati, Phys. Lett. B 265, 258 (1991).

\refis{tvmb} T. Vachaspati and M. Barriola, Phys. Rev. Lett. 69,
1867 (1992).

\refis{mbtvmb} M. Barriola, T. Vachaspati and M. Bucher,
``Embedded Defects'', Phys. Rev. D, to be published; TUTP-93-7.

\refis{marty} M. B. Einhorn and R. Savit, Phys. Lett. B 77, 295 (1978).

\refis{minos} M. Axenides and A. Johansen, NBI-HE-93-74.

\refis{kunz} J. Kunz, B. Kleihaus and Y. Brihaye, Phys. Rev. D 46,
3587 (1992).

\refis{efarhi} E. Farhi, V. V. Khoze and R. Singleton Jr.,
Phys. Rev. D 47, 5551 (1993).

\refis{wp} W. Perkins, Phys. Rev. D 47, R5224 (1993).

\refis{japo} For a review see J. Ambjorn and  P. Olesen,
Int. J. Mod. Phys. A 5, 4525 (1990).

\refis{hnpo} H. B. Nielsen and P. Olesen, Nucl. Phys. B 61, 45 (1973).

\refis{poprivate} P. Olesen (private communication).

\refis{fkpo} F. R. Klinkhamer and P. Olesen, NIKHEF-H/94-02.

\refis{james1} M. James, DAMTP-HEP-94-13.

\refis{james2} M. James (private communication).

\refis{memj} M. A. Earnshaw and M. James, hep-ph/9308223 (1993).

\refis{mhmj} M. Hindmarsh and M. James, hep-ph/9307205 (1993).

\refis{gdgs} G. Dvali and G. Senjanovic, Phys. Rev. Lett. 71, 2376 (1993).

\refis{nambu} Y. Nambu, Nucl. Phys. B130, 505 (1977).

\refis{nm} N. S. Manton, Phys. Rev. D28, 2019 (1983).

\refis{fknm} F. R. Klinkhamer and N. S. Manton, Phys. Rev. D30, (1984)
2212.

\refis{tv1}  T. Vachaspati, Phys. Rev. Lett. 68, 1977 (1992);
69, 216(E) (1992); Nucl. Phys. B397, 648 (1993).

\refis{mbgf} J. J. Moreau, C. R. Acad. Sci. Paris 252, 2810 (1961);
H. K. Moffat, J. Fluid Mech. 35, 117 (1969); M. Berger and G. Field,
J. Fluid Mech. 147, 133 (1984).

\refis{mjlptv} M. James, L. Perivolaropoulos and T. Vachaspati, Phys.
Rev. D46 (1992) R5232; Nucl. Phys. B395, 534 (1993).

\refis{plrm} P. Laguna-Castillo and R. A. Matzner,
Phys. Rev. D 36, 3663 (1987).

\refis{flanders} For example see H. Flanders, ``Differential Forms''
(Academic Press, 1963).

\refis{knots} For example, see L. H. Kauffman, ``On Knots'', Princeton
University Press (1987).

\refis{misha} M. E. Shaposhnikov, JETP Lett. 44, 465 (1986).

\refis{tvavformation} T. Vachaspati and A. Vilenkin,
Phys. Rev. D 30, 2036 (1984).

\endreferences



\endjnl
\end